\documentstyle[prd,aps]{revtex}
\begin{document}

\title{Exact path integral of the hydrogen atom and 
the Jacobi's principle of least action\footnote{UT-756,1996.  Talk presented at 
Inauguration Conference of Asia Pacific Center for Theoretical Physics,  Seoul, Korea, June 4-10,  1996. To be published in the Proceedings(World Scientific, Singapore) }}

\author{ Kazuo Fujikawa}
\address{
Department of Physics, University of Tokyo\\
         Bunkyo-ku, Tokyo  113, Japan}
\maketitle
\large
\begin{abstract}
The general treatment of a separable Hamiltonian of Liouville-type is 
well-known in  operator formalism. A path integral counterpart is 
formulated if one starts with the Jacobi's principle of least action, and 
a path integral evaluation of the Green's function for the hydrogen atom by Duru and Kleinert is recognized as a special case. The Jacobi's principle of least action for given energy is reparametrization invariant, and the  
separation of variables in operator formalism corresponds to a choice of
gauge in path integral. The Green's function is shown to be gauge 
independent,if the operator ordering is properly taken into account.  These properties are illustrated by evaluating an  exact path integral of the 
Green's function for the hydrogen atom in parabolic coordinates.
\end{abstract}

\section{Introduction}
The path 
integral treatment of the hydrogen atom is interesting not only for a 
methodological interest but also for a pedagogical purpose.  
After several early attempts[1,2], Duru and Kleinert[3] showed an elegant  
path integral method to evaluate the Green's function for the hydrogen atom 
exactly. Two basic ingredients in their method are the use of a re-scaled time variable and the  so called  Kustaanheimo-Stiefel transformation[4]which reveals the O(4) symmetry explicitly in the coordinate space. Many of the clarifying works of this approach have been published [5] - [14]. 
It has been shown elsewhere[15] that this problem is treated 
in a more general setting of gauge theory if one formulates the problem  
on the basis of the Jacobi's principle of least action, which is reparametrization invariant.

The general technique of gauge theory is thus applicable to the evaluation of 
path integral, and  a suitable choice of gauge simplifies the problem such as 
the hydrogen atom. In particular, the Green's function is  shown to be gauge independent. The use of  the Kustaanheimo-Stiefel transformation is  rather 
technical and it is not essential in solving the problem exactly. We in fact 
show a simple trick in parabolic coordinates[16] which solves the hydrogen
atom exactly.

\section{Hydrogen Atom}

{\bf 2.1, Analysis in Parabolic Coordinates}

We analyze the hydrogen atom by starting  with the Hamiltonian written in terms of   parabolic coordinates
\begin{equation}
H(\xi, \eta, \varphi)  = \frac{1}{2m(\xi + \eta)}(\xi p_{\xi}^{2} + \eta p_{\eta}^{2}) + \frac{1}
{8m\xi \eta}p_{\varphi}^{2} - \frac{e^{2}}{\xi + \eta}
\end{equation}
where the parabolic coordinates are introduced by
$\xi = \frac{1}{2}(r - z),\eta =\frac{1}{2}(r + z)$
and $\varphi$ stands for the azimuthal angle around the $z$ axis. We further 
perform a canonical transformation which simplifies the kinetic term in $H$
as
\begin{eqnarray}
\xi &=& \frac{1}{4}u^{2}\ \ \ \ \ \ \ \ \ ,0\leq u < \infty \nonumber\\
\sqrt{\xi}p_{\xi}&=& p_{u}\nonumber\\
\eta &=& \frac{1}{4}v^{2}\ \ \ \ \ \ \ \ \ ,0\leq v < \infty \nonumber\\
\sqrt{\eta}p_{\eta}&=& p_{v}
\end{eqnarray}
and the Hamiltonian becomes
\begin{equation}
H = \frac{1}{2m}(\frac{4}{u^{2} + v^{2}})[p_{u}^{2} + \frac{1}{u^{2}}p_{\varphi}^{2} + p_{v}^{2} + \frac{1}{v^{2}}p_{\varphi}^{2}] 
- \frac{4}{u^{2} + v^{2}}e^{2}
\end{equation}
where $ r = \xi + \eta = (u^{2} + v^{2})/4$. This Hamiltonian is not yet a separable one of Liouville-type.

One may solve the Schroedinger equation 
\begin{equation}
\hat{H} \psi = E\psi
\end{equation}
or equivalently
\begin{equation}
\hat{H}_{T}\psi = 0
\end{equation}
with
\begin{equation}
\hat{H}_{T} = \frac{1}{2m}[{\hat{p}_{u}}^{2} + \frac{1}{u^{2}}{\hat{p}_{\varphi}}^{2} + {\hat{p}_{v}}^{2}+  \frac{1}{v^{2}}{\hat{p}_{\varphi}}^{2}] -e^{2} 
+ \frac{m\omega^{2}}{2}(u^{2} +v^{2})
\end{equation}
where $\omega$ is defined by
\begin{equation}
\frac{1}{2}m{\omega}^{2} = -\frac{1}{4}E
\end{equation}
We consider the case $E < 0$ for the moment.
$\hat{H}_{T}$ stands for the total Hamiltonian defined by a specific gauge 
condition; a  general definition of $\hat{H}_{T}$ will
be given later in (39).  

Eq.(5) may be rewritten in an equivalent form as
\begin{eqnarray}
\hat{\tilde{H}}_{T}\psi &=& 0\nonumber\\
(\hat{p}_{\varphi} -  \hat{p}_{{\varphi}^{\prime}})\psi& =& 0
\end{eqnarray}
We here introduced auxiliary variables  $(\hat{p}_{{\varphi}^{\prime}},
\varphi^{\prime})$ as
\begin{eqnarray}
\hat{\tilde{H}}_{T}&=& \frac{1}{2m}[\hat{p}_{u}^{2} + \frac{1}{u^{2}}\hat{p}_{\varphi}^{2} + \hat{p}_{v}^{2} +  \frac{1}{v^{2}}\hat{p}_{{\varphi}^{\prime}}]  + \frac{m\omega^{2}}{2}(u^{2} +v^{2}) - e^{2}\nonumber\\
&=&\frac{1}{2m}\vec{p}_{u}^{2} + \frac{m\omega^{2}}{2}\vec{u}^{2} +
   \frac{1}{2m}\vec{p}_{v}^{2} + \frac{m\omega^{2}}{2}\vec{v}^{2} - e^{2}
\end{eqnarray}
and we  defined 
\begin{eqnarray}
\vec{u} &=& ( u_{1},u_{2})= (u\cos\varphi, u\sin\varphi)\nonumber\\
\vec{p}_{u}^{2} &=& \hat{p}_{u}^{2} + \frac{1}{u^{2}}\hat{p}_{\varphi}^{2}\nonumber\\
\vec{v} &=& ( v_{1},v_{2})= (v\cos{\varphi^{\prime}},v\sin{\varphi^{\prime}})\nonumber\\
\vec{p}_{v}^{2} &=& \hat{p}_{v}^{2} + \frac{1}{v^{2}}\hat{p}_{{\varphi}^{\prime}}^{2}
\end{eqnarray} 
The subsidiary condition in (8) replaces the use of the Kustaanheimo-Stiefel transformation, and at the same time it renders a Hamiltonian of Liouville-type. This introduction of auxiliary variables (10) has been discussed by Ravndal and Toyoda[16]. 

A general procedure to deal with a completely separated operator such as $\hat{\tilde{H}}_{T}$ in (9) is to consider an evolution operator for a parameter
$\tau$ defined by
\begin{eqnarray}
&&\langle \vec{u}_{b}, \vec{v}_{b}|e^{-i\hat{\tilde{H}}_{T}\tau/\hbar}|\vec{u}_{a}, \vec{v}_{a}\rangle\nonumber\\
&&= e^{ie^{2}\tau}\langle \vec{u}_{b}|exp[ -(i/\hbar)(\frac{1}{2m}\vec{p}_{u}^{2} + \frac{m\omega^{2}}{2}\vec{u}^{2})\tau]|\vec{u}_{a}\rangle\nonumber\\
&&\times \langle \vec{v}_{b}|exp[ -(i/\hbar)(\frac{1}{2m}\vec{p}_{v}^{2} + \frac{m\omega^{2}}{2}\vec{v}^{2})\tau]|\vec{v}_{a}\rangle\nonumber\\
&&= e^{ie^{2}\tau}(\frac{m\omega}{2\pi i\hbar \sin \omega \tau})^{4/2}\nonumber\\
&&\times exp\{\frac{im\omega}{2\hbar \sin \omega\tau}[(\vec{u}_{b}^{2} +\vec{v}_{b}^{2} + \vec{u}_{a}^{2} + \vec{v}_{a}^{2})\cos \omega\tau -2\vec{u}_{b}\vec{u}_{a} -
2\vec{v}_{b}\vec{v}_{a}]\}
\end{eqnarray}
where we used the exact result for a simple harmonic oscillator[17].

A crucial observation here is that $\hat{p}_{\varphi}$ and $\hat{p}_{{\varphi}^{\prime}}$ are preserved during the evolution dictated by the operator $\hat{\tilde{H}}_{T}$ in (9), since $ [\hat{p}_{\varphi}, \hat{\tilde{H}}_{T}] =  
[\hat{p}_{{\varphi}^{\prime}}, \hat{\tilde{H}}_{T}] =  0$. It is then sufficient to impose the constraint (8) only on the initial state , for example. 
Starting with a general state belonging to the eigenvalues $\hat{p}_{\varphi}=
m$ and $\hat{p}_{{\varphi}^{\prime}} = m^{\prime}$ 
\begin{equation}
e^{i m\varphi} e^{i m^{\prime}\varphi^{\prime}} 
\end{equation}
we can use the following trick 
\begin{equation}
\int_{0}^{2\pi} \frac{d\theta}{2\pi} e^{i m(\varphi + \theta)}e^{i m^{\prime}
(\varphi^{\prime} - \theta)} = \delta_{m,m^{\prime}}e^{i m(\varphi + \varphi^{\prime})}
\end{equation}
to project out the state satisfying $\hat{p}_{\varphi}= \hat{p}_{{\varphi}^{\prime}}$, and $\varphi + \varphi^{\prime}$ is regarded as the actual azimuthal
angle.

We thus obtain
\begin{eqnarray}
&&\langle u_{b}, v_{b}, (\varphi + \varphi^{\prime})_{b}|e^{-i \hat{H}_{T}\tau/\hbar} | u_{a}, v_{a}, (\varphi + \varphi^{\prime})_{a}\rangle\nonumber\\
&&= e^{ie^{2}\tau}(\frac{m\omega}{2\pi i\hbar \sin \omega \tau})^{2}
\int_{0}^{2\pi}\frac{d\theta}{2\pi}exp\{\frac{im\omega}{2\hbar \sin \omega\tau}[(\vec{u}_{b}^{2} +\vec{v}_{b}^{2} + \vec{u}_{a}^{2} + \vec{v}_{a}^{2})\cos \omega\tau -2\vec{u}_{b}\vec{u}_{a} -
2\vec{v}_{b}\vec{v}_{a}]\}\nonumber\\
&&= e^{ie^{2}\tau}(\frac{m\omega}{2\pi i\hbar \sin \omega \tau})^{2}\times \nonumber\\
&&\int_{0}^{2\pi}\frac{d\theta}{2\pi}exp\{\frac{im\omega}{2\hbar \sin \omega\tau}[4(\xi_{a} + \xi_{b} + \eta_{a} +
\eta_{b} )\cos \omega\tau - 4\sqrt{2}(r_{a}r_{b} + \vec{x}_{a}\vec{x}_{b})^{1/2}\cos (\theta + \gamma) ]\}\nonumber\\
&&=  e^{ie^{2}\tau}(\frac{m\omega}{2\pi i\hbar \sin \omega \tau})^{2}
exp\{\frac{2 im\omega}{\hbar \sin \omega\tau}
(r_{a} + r_{b})\cos \omega\tau\}
I_{0}(\frac{2\sqrt{2} i m\omega}{\hbar \sin \omega \tau}(r_{a}r_{b} + \vec{x}_{a}\vec{x}_{b})^{1/2})\nonumber\\
&& 
\end{eqnarray}
In this evaluation we start with the relation
\begin{equation}
\vec{u}_{b}\vec{u}_{a} + \vec{v}_{b}\vec{v}_{a}
= u_{b}u_{a}\cos \Delta\varphi + v_{b}v_{a}\cos \Delta\varphi^{\prime}
\end{equation}
with $\Delta\varphi = \varphi_{b} - \varphi_{a}, \Delta\varphi^{\prime} =
\varphi^{\prime}_{b} - \varphi^{\prime}_{a}$, and  
\begin{eqnarray}
&&u_{b}u_{a}\cos (\Delta\varphi+\theta) + v_{b}v_{a}\cos (\Delta\varphi^{\prime}-\theta)\nonumber\\
&&= (u_{b}u_{a}\cos \Delta\varphi + v_{b}v_{a}\cos \Delta\varphi^{\prime})
\cos\theta \nonumber\\
&&+(- u_{b}u_{a}\sin \Delta\varphi + v_{b}v_{a}\sin \Delta\varphi^{\prime})
\sin\theta \nonumber\\ 
&&=4\sqrt{\xi_{b}\xi_{a} +\eta_{b}\eta_{a} +2\sqrt{\xi_{b}\xi_{a}\eta_{b}\eta_{a}}\cos(\Delta\varphi + \Delta\varphi^{\prime})}\cos(\theta +\gamma)\nonumber\\
&&=2\sqrt{2}\sqrt{r_{a}r_{b} + z_{a}z_{b} + \rho_{a}\rho_{b}\cos(\Delta\varphi + \Delta\varphi^{\prime})}\cos(\theta +\gamma)\nonumber\\
&&=2\sqrt{2}\sqrt{r_{a}r_{b} + \vec{x}_{a}\vec{x}_{b}}\cos(\theta +\gamma)
\end{eqnarray}
where we used the definition of  variables in (2),  and $\gamma$ is a number independent of $\theta$. We also defined a modified Bessel function
\begin{equation}
I_{0}(\frac{2\sqrt{2} i m\omega}{\hbar \sin \omega \tau}(r_{a}r_{b} + \vec{x}_{a}\vec{x}_{b})^{1/2}) = \int_{0}^{2\pi}\frac{d\theta}{2\pi}exp\{\frac{2\sqrt{2}im\omega (r_{a}r_{b} + \vec{x}_{a}\vec{x}_{b})^{1/2}   }{\hbar \sin \omega\tau}\cos \theta  \}
\end{equation}

The parameter $\tau$ is arbitrary, and we eliminate $\tau$ to obtain a physically meaningful quantity by 
\begin{eqnarray}
&&i\int_{0}^{\infty} d\tau \langle u_{b}, v_{b}, (\varphi + \varphi^{\prime})_{b}|e^{-i \hat{H}_{T}\tau/\hbar} | u_{a}, v_{a}, (\varphi + \varphi^{\prime})_{a}\rangle\nonumber\\
&&=\langle u_{b}, v_{b}, (\varphi + \varphi^{\prime})_{b}|\frac{\hbar}{\hat{H}_{T}} | u_{a}, v_{a}, (\varphi + \varphi^{\prime})_{a}\rangle\nonumber\\
&&= i\int_{0}^{\infty} d\tau 
e^{ie^{2}\tau}(\frac{m\omega}{2\pi i\hbar \sin \omega \tau})^{2}
exp\{\frac{2 im\omega}{\hbar \sin \omega\tau}
(r_{a} + r_{b})\cos \omega\tau\}\nonumber\\
&&\times I_{0}(\frac{2\sqrt{2} i m\omega}{\hbar \sin \omega \tau}(r_{a}r_{b} + \vec{x}_{a}\vec{x}_{b})^{1/2})\nonumber\\
&&=\frac{m\omega}{2\pi^{2}\hbar^{2}}\int_{0}^{1}d\lambda \lambda^{-\nu}
\frac{1}{(1-\lambda)^{2}}
exp [ \frac{-2m\omega}{\hbar}(r_{a} + r_{b})(\frac{1+\lambda}{1 - \lambda})]I_{0}(\frac{4\sqrt{2}m\omega}{\hbar}
\frac{\lambda^{1/2}
}{1-\lambda}(r_{a}r_{b} + \vec{x}_{a}\vec{x}_{b})^{1/2})
\nonumber\\
&&
\end{eqnarray}
where we rotated $\tau$ by 90 degrees , $\tau \rightarrow -i\tau$, and defined
\begin{eqnarray}
\lambda &=& e^{-2\omega \tau}\nonumber\\
\nu &=& e^{2}/2\omega
\end{eqnarray}

We next show that (18) gives an exact Green's function for the hydrogen atom by noting the sequence
\begin{eqnarray}
&&\langle u_{b}, v_{b}, \varphi_{b}|\frac{\hbar}{\hat{H}_{T}} | u_{a}, v_{a}, 
\varphi_{a}\rangle\nonumber\\
&&=\langle \xi_{b}, \eta_{b}, \varphi_{b}|\frac{\hbar}{(\frac{1}{\hat{\xi} + \hat{\eta}})\hat{H}_{T}(\xi, \eta, \varphi)} | \xi_{a}, \eta_{a}, 
\varphi_{a}\rangle (\frac{1}{\xi_{a} + \eta_{a}})\nonumber\\
&&=\langle \xi_{b}, \eta_{b}, \varphi_{b}|\frac{\hbar}{\hat{H}(\xi, \eta, \varphi) - E} | \xi_{a}, \eta_{a}, 
\varphi_{a}\rangle (\frac{1}{\xi_{a} + \eta_{a}})\nonumber\\
&&=\frac{\hbar}{\hat{H}(\xi_{b},\eta_{b},\varphi_{b}) - E}(\frac{1}{\sqrt{\xi_{b}+ \eta_{b}}}\langle \xi_{b}, \eta_{b}, \varphi_{b}| \xi_{a}, \eta_{a}, 
\varphi_{a}\rangle \frac{1}{\sqrt{\xi_{a} + \eta_{a}}})\nonumber\\
&&= \frac{1}{4\pi}\langle \vec{x}_{b}|\frac{\hbar}{\hat{\vec{p}}^{2}/{2m}
- e^{2}/r - E}|\vec{x}_{a}\rangle
\end{eqnarray}
where we used $\varphi$ in place of $\varphi + \varphi^{\prime}$ and the relation $(\hat{A}\hat{B})^{-1} = \hat{B}^{-1}\hat{A}^{-1}$.

The volume element changes in this transition from $\hat{H}_{T}$ to $\hat{H}$ as
\begin{eqnarray}
dV_{0} &=& 2\pi uvdudvd\varphi\nonumber\\
\rightarrow  dV &=& (\xi + \eta)dV_{0} = 4\pi\times 2(\xi + \eta)d\xi d\eta d\varphi\nonumber\\
   &=& 4\pi\times  r^{2}dr d\cos \theta d\varphi
\end{eqnarray}
The bra- and ket- vectors in (20) are normalized in the combination 
\begin{eqnarray}
\int dV_{0}| u, v, \varphi\rangle \langle u, v, \varphi| &=& 1\nonumber\\
\int dV| \xi, \eta, \varphi\rangle \frac{1}{\xi + \eta} \langle \xi, \eta, \varphi| &=& 1\nonumber\\
\int d^{3}x |\vec{x}\rangle \langle \vec{x}| &=& 1
\end{eqnarray}
and the extra factor of $4\pi$ in $dV = 4\pi r^{2}dr d\cos \theta d\varphi$ requires
the appearance of the factor of $1/4\pi$ in the last expression  in (20). The appearance of $2\pi$ in $dV_{0}$ is an artifact of the variable $\varphi^{\prime}$ in (10). This normalization condition of bra-  and ket- vectors together 
with a symmetry in $\vec{x}_{a}$ and $\vec{x}_{b}$ justify the identification
(20). A more explicit and concrete analysis of eqs.(20)$\sim $
 (22) will be given in connection with the Jacobi's principle later.

As for the operator ordering, the momentum operator  changes in (20) as
\begin{eqnarray}
\hat{p}^{2}_{u} + \hat{p}^{2}_{v} &=& (\frac{\hbar}{i})^{2}[\frac{1}{u}\partial
_{u}u\partial_{u} + \frac{1}{v}\partial_{v}v\partial_{v}]\nonumber\\
&=&(\frac{\hbar}{i})^{2}[\partial_{\xi}\xi\partial_{\xi} + \partial_{\eta}\eta\partial_{\eta}]\nonumber\\ 
&=& \hat{p}_{\xi}\xi\hat{p}_{\xi} + \hat{p}_{\eta}\eta\hat{p}_{\eta}
\end{eqnarray}
and 
\begin{equation}
(\frac{1}{\xi + \eta})( \hat{p}_{\xi}\xi \hat{p}_{\xi} + \hat{p}_{\eta}\eta \hat{p}_{\eta} ) + \frac{1}{4\xi\eta}{\hat{p}_{\varphi}}^{2} ={ \hat{\vec{p}}}^{2}
\end{equation}
where the right-hand side is written in cartesian coordinates.
We note that $dV_{0}$ and $dV$ in (21) respectively render $\hat{H}_{T}$ and 
$\hat{H}(\xi, \eta, \varphi )$ hermitian.

Combining (18),(20) and (24), we have thus established the exact Green's 
function including the {\em  operator ordering} 
\begin{eqnarray}
\langle \vec{x}_{b}|\frac{\hbar}{\hat{\vec{p}}^{2}/{2m}
- e^{2}/r - E}|\vec{x}_{a}\rangle &=& 
\frac{2 m^{2}\omega}{\pi \hbar^{2}}\int_{0}^{1}d\lambda \lambda^{-\nu}
\frac{1}{(1-\lambda)^{2}}
exp [ \frac{-2m\omega}{\hbar}(r_{a} + r_{b})(\frac{1+\lambda}{1 - \lambda})]
\nonumber\\
&& \times I_{0}(\frac{4\sqrt{2}m\omega}{\hbar}
\frac{\lambda^{1/2}}{1-\lambda}(r_{a}r_{b} + \vec{x}_{a}\vec{x}_{b})^{1/2})
\end{eqnarray}
It is known that this formula, which was first derived by Duru and 
Kleinert[3], is a Fourier transform of Schwinger's momentum space representation[18]. The continuation to the scattering problem with $E > 0$ is performed by
the replacement
\begin{equation}
\omega \rightarrow (-i)\omega, \ \ \nu \rightarrow i\nu
\end{equation}
in the above formula.

One can understand the spectrum of the hydrogen atom by looking at $\hat{\tilde{H}}_{T}$ in (9)[16].
If one defines the oscillator variables 
\begin{eqnarray}
a_{k} &=& \frac{1}{\sqrt{2}}[\sqrt{\frac{m\omega}{\hbar}} u_{k} + 
          \frac{i}{\sqrt{m\omega \hbar}}\hat{p}_{u_{k}}], \nonumber\\
\tilde{a}_{k} &=& \frac{1}{\sqrt{2}}[\sqrt{\frac{m\omega}{\hbar}} v_{k} + 
          \frac{i}{\sqrt{m\omega \hbar}}\hat{p}_{v_{k}}],\ \ k = 1, 2
\end{eqnarray}
one obtains 
\begin{eqnarray}
\hat{\tilde{H}}_{T} &=& \hbar\omega [\sum_{k=1}^{2}( a_{k}^{\dagger}a_{k} 
            + \tilde{a}_{k}^{\dagger}\tilde{a}_{k}) + 2 ] - e^{2}\nonumber\\
\hat{p}_{\varphi} &=& i\hbar [a_{1}^{\dagger}a_{2} - a_{2}^{\dagger}a_{1}]
\nonumber\\
\hat{p}_{\varphi^{\prime}} &=& i\hbar [\tilde{a}_{1}^{\dagger}\tilde{a}_{2} - 
\tilde{a}_{2}^{\dagger}\tilde{a}_{1}]
\end{eqnarray}
After a unitary transformation
\begin{eqnarray}
a_{1} &=& \frac{1}{\sqrt{2}}( b_{1} - i b_{2})\nonumber\\
a_{2} &=& \frac{1}{\sqrt{2}}( - i b_{1} +  b_{2})
\end{eqnarray}
and a similar transformation of $\tilde{a}_{1}$ and $\tilde{a}_{2}$, one 
obtains
\begin{eqnarray}
\hat{\tilde{H}}_{T} &=& \hbar\omega [\sum_{k=1}^{2}( b_{k}^{\dagger}b_{k} 
            + \tilde{b}_{k}^{\dagger}\tilde{b}_{k}) + 2 ] - e^{2}\nonumber\\
\hat{p}_{\varphi} &=& \hbar [b_{1}^{\dagger}b_{1} - b_{2}^{\dagger}b_{2}]
\nonumber\\
\hat{p}_{\varphi^{\prime}} &=& \hbar [\tilde{b}_{1}^{\dagger}\tilde{b}_{1} - 
\tilde{b}_{2}^{\dagger}\tilde{b}_{2}]
\end{eqnarray}
By defining the number operators
\begin{eqnarray}
n_{k} &=& b^{\dagger}_{k}b_{k},\nonumber\\
\tilde{n}_{k} &=& \tilde{b}^{\dagger}_{k}\tilde{b}_{k}, \ \ k = 1, 2
\end{eqnarray}
the total Hamiltonian is given by 
\begin{eqnarray}
\hat{\tilde{H}}_{T} &=& \hbar\omega [n_{1} + n_{2} + \tilde{n}_{1} + \tilde{n}_{2} + 2 ] - e^{2}\nonumber\\
&=& \hbar\omega [ 2n_{1} - \hat{p}_{\varphi}/\hbar + 2\tilde{n}_{2} + \hat{p}_{\varphi^{\prime}}/\hbar + 2 ] - e^{2}\nonumber\\
&=& 2\hbar\omega [ n_{1} + \tilde{n}_{2} + 1 ] - e^{2} 
\end{eqnarray}
by noting the physical state condition $\hat{p}_{\varphi}= \hat{p}_{\varphi^{\prime}}$.

We thus define the principal quantum number (or its operator) $n$ by 
\begin{equation}
n = n_{1} + \tilde{n}_{2} + 1 = 1, 2, 3, .....
\end{equation}
and the physical state condition 
\begin{equation}
( 2 n \hbar\omega - e^{2})\psi_{phys} = 0
\end{equation}
gives rise to the Bohr spectrum
\begin{equation}
E = - \frac{mc^{2}}{2} (\frac{e^{2}}{\hbar c})^{2}\frac{1}{n^{2}}, \ \ n= 1, 2, 3,...
\end{equation}
by noting the definition of $\omega$ in (7). As for the degeneracy of  states with a fixed $n$, it is confirmed that each
state with a given $n$ has $n^{2}$ degeneracy.

{\bf 2.2, Jacobi's Principle of Least Action}

The meaning of the total Hamiltonian in (6) becomes transparent , if one starts 
with a Nambu-Goto-type Lagrangian ( the Jacobi's principle of least  action for 
a given E ) which is reparametrization invariant, 
\begin{eqnarray}
S &=& \int_{0}^{\tau} L d\tau = \int_{0}^{\tau} d\tau \sqrt{2m(E-V(r))(\frac{d\vec{x}}{d\tau})^{2}}\nonumber\\
&=& \int \sqrt{2m(E-V(r))(d\vec{x})^{2}}
\end{eqnarray}
The momenta conjugate to coordinates are  then defined by
\begin{equation}
\vec{p} = \frac{\partial L}{\partial \dot{\vec{x}}} = \sqrt{2m(E-V(r))}(\frac{d\vec{x}}{d\tau})/\sqrt{(\frac{d\vec{x}}{d\tau})^{2}}
\end{equation}
and  one obtains a vanishing Hamiltonian as a result of reparametrization
invariance and a first-class constraint $\phi$, which is the generator of 
reparametrization gauge symmetry,
\begin{eqnarray}
H &=& \vec{p}\dot{\vec{x}} - L = 0\nonumber\\
\phi (\vec{x}, \vec{p})& =& \frac{\vec{p}^{2}}{2m} +
 V(r) - E \simeq 0
\end{eqnarray}
Following Dirac[19], the total Hamiltonian is defined by
\begin{eqnarray}
H_{T}& =&  H + \alpha (\vec{x},\vec{p})\phi (\vec{x}, \vec{p})\nonumber\\
     & =&  \alpha (\vec{x},\vec{p})\phi (\vec{x}, \vec{p})\simeq 0
\end{eqnarray}
and the  function $\alpha (\vec{x}, \vec{p})$ specifies a choice of 
gauge and fixes the arbitrary parameter $\tau$ in (36), which parametrizes 
the orbit  for a given $E$. A change of the parameter $\tau$ to 
$\tau - \delta\beta (\tau, \vec{x}, \vec{p})$ is generated by 
$\delta\beta (\tau, \vec{x}, \vec{p})\alpha (\vec{x},\vec{p})\phi (\vec{x}, \vec{p}) = \delta\beta (\tau, \vec{x}, \vec{p})H_{T}$, for example,
\begin{eqnarray}
\delta\vec{x}(\tau) &=& \vec{x}^{\prime}(\tau) -  \vec{x}(\tau)\nonumber\\
&=& \vec{x}(\tau + \delta\beta) - \vec{x}(\tau)\nonumber\\
&=& \{ \vec{x}, \delta\beta H_{T}\}_{PB}\nonumber\\
&=& \delta\beta (\tau, \vec{x}, \vec{p})\frac{d}{d\tau}\vec{x}(\tau)
\end{eqnarray}
in terms of the Poisson bracket, since $\vec{x}^{\prime}(\tau - \delta\beta) =
\vec{x}(\tau)$.

Quantization is performed by
\begin{equation}
i\hbar\frac{\partial}{\partial\tau}\psi = \hat{H}_{T}\psi  
\end{equation}
with a physical state condition
\begin{equation}
\hat{\alpha}(\vec{x}, \vec{p})\hat{\phi} (\vec{x}, \vec{p})
\psi_{phy} = 0
\end{equation}
A specific choice of the gauge $\alpha (\vec{x}, \vec{p}) = r = \xi + \eta$ 
leads to the  Hamiltonian $\hat{H}_{T}$ in (6) and the choice $\alpha (\vec{x}, \vec{p}) = 1$ gives the original static Schroedinger equation(4), since the 
states $\psi$ in (5) and (4) are physical states. Eq.(41) gives rise to 
the evolution operator in (11). 

We now explain the relations (20)$\sim$(22) in a more concrete manner. We 
start with eq.(20) for a generic negative $E$
\begin{eqnarray}
G(E; \xi_{b},\eta_{b},\varphi_{b};\xi_{a},\eta_{a},\varphi_{a})
&=& \langle \xi_{b},\eta_{b},\varphi_{b}|\frac{\hbar}{
\hat{H}_{T}(E; \xi, \eta, \varphi)}|\xi_{a},\eta_{a},\varphi_{a}\rangle\nonumber\\
&\equiv& \sum_{n} \phi_{n}(E; \xi_{b},\eta_{b},\varphi_{b})\frac{\hbar}{\lambda_{n}(E)}\phi_{n}^{\star}(E; \xi_{a},\eta_{a},\varphi_{a})
\end{eqnarray}
with
\begin{eqnarray}
\hat{H}_{T}(E; \xi, \eta, \varphi)\phi_{n}(E; \xi,\eta,\varphi )
&=& \lambda_{n}(E)\phi_{n}(E; \xi,\eta,\varphi )\nonumber\\
\int \phi_{n}^{\star}(E; \xi,\eta,\varphi )\phi_{l}(E; \xi,\eta,\varphi ) dV_{0} &=& \delta_{n, l}\nonumber\\
\lambda_{n}(E) = 2n\hbar \sqrt{- \frac{E}{2m}} - e^{2}\nonumber\\
dV_{0} = 4\pi\times 2d\xi d\eta d\varphi
\end{eqnarray}
where we used the result in (32) and also the variables $(\xi, \eta, \varphi )$ instead of $(u, v, \varphi )$ for notational simplicity. The summation over $n$ in (43) is formal including the $n^{2}$ degeneracy. Note that the complete orthonormal states $\{ \phi_{n}\}$ in (43) are all {\em unphysical} off-
shell states. In path integral, the summation in (43) is exactly evaluated in
(18). 

We next rewrite $G(E; \xi_{b},\eta_{b},\varphi_{b};\xi_{a},\eta_{a},\varphi_{a})$ in terms of physical on-shell states by writing an unsubtracted dispersion
relation (i.e., paying attention only to the pole structure in $E$) as
\begin{equation}
G(E; \xi_{b},\eta_{b},\varphi_{b};\xi_{a},\eta_{a},\varphi_{a}) =
\sum_{n} \phi_{n}(E_{n}; \xi_{b},\eta_{b},\varphi_{b})\frac{\hbar}{(E_{n} - E)(- 
\frac{\partial \lambda_{n}(E_{n})}{\partial E_{n}})}\phi_{n}^{\star}(E_{n}; \xi_{a},\eta_{a},\varphi_{a})
\end{equation}
by noting 
\begin{equation}
\lambda_{n}(E) = \lambda_{n}(E_{n}) + (E-E_{n})\frac{\partial \lambda_{n}(E_{n})}{\partial E_{n}} = (E-E_{n})(\frac{-e^{2}}{2E_{n}})
\end{equation}
for $E \approx E_{n}$.

When one defines 
\begin{equation}
\psi_{n}(E_{n}; \xi,\eta,\varphi ) = \frac{1}{\sqrt{-\frac{\partial \lambda_{n}(E_{n})}{\partial E_{n}}}}\phi_{n}(E_{n}; \xi,\eta,\varphi )
\end{equation}
one can show the orthonormality relations of physical on-shell states
\begin{eqnarray}
\int \psi_{n}^{\star}(E_{n}; \xi,\eta,\varphi )\psi_{l}(E_{l}; \xi,\eta,\varphi )(\xi + \eta )dV_{0} &=&
\int \psi_{n}^{\star}(E_{n}; \xi,\eta,\varphi )\psi_{l}(E_{l}; \xi,\eta,\varphi )dV \nonumber\\
&=& \delta_{n,l}
\end{eqnarray}
with $dV = (\xi + \eta )dV_{0}$.
First of all, from the physical state condition 
\begin{eqnarray}
&&\hat{H}_{T}(E_{n}; \xi, \eta, \varphi )\phi_{n} (E_{n}; \xi, \eta, \varphi )
\\
&=&\{ \frac{1}{2m}(\hat{p}_{\xi} \xi \hat{p}_{\xi} + \hat{p}_{\eta}  \eta \hat{p}_{\eta}) + \frac{1}{8m}(\frac{1}{\xi} + \frac{1}{\eta})\hat{p}_{\varphi}^{2}  - E_{n}(\xi + \eta ) - e^{2} \}\phi_{n} (E_{n}; \xi, \eta, \varphi ) = 0\nonumber
\end{eqnarray}
one can establish the orthogonality relation
\begin{equation}
(E_{n} - E_{l})\int \phi_{n}^{\star}(E_{n}; \xi,\eta,\varphi )\phi_{l}(E_{l}; \xi,\eta,\varphi )(\xi + \eta )dV_{0} = 0
\end{equation}
for $n \neq l$. Also from the relation (49) and the fact that the ``eigenvalue'' $e^{2}$ is equally distributed for the kinetic and potential terms for
harmonic oscillators(in terms of $u$ and $v$ variables) , we have
\begin{equation}
- E_{n}\int \phi_{n}^{\star}(E_{n}; \xi,\eta,\varphi )\phi_{n}(E_{n}; \xi,\eta,\varphi )(\xi + \eta )dV_{0} = \frac{e^{2}}{2}
\end{equation}
namely
\begin{equation}
\frac{1}{-\frac{\partial \lambda_{n}(E_{n})}{\partial E_{n}}}\int \phi_{n}^{\star}(E_{n}; \xi,\eta,\varphi )\phi_{n}(E_{n}; \xi,\eta,\varphi )(\xi + \eta )dV_{0} = 1
\end{equation}
by noting (44). This proves (48).

From (45), we finally arrive at the expression
\begin{eqnarray}
G(E; \xi_{b},\eta_{b},\varphi_{b};\xi_{a},\eta_{a},\varphi_{a}) &=&
\sum_{n} \psi_{n}(E_{n}; \xi_{b},\eta_{b},\varphi_{b})\frac{\hbar}{E_{n} - E}
\psi_{n}^{\star}(E_{n}; \xi_{a},\eta_{a},\varphi_{a})\nonumber\\
&=& \langle \xi_{b},\eta_{b},\varphi_{b}|\frac{\hbar}{\hat{H}(\xi,\eta,\varphi ) - E}|\xi_{a},\eta_{a},\varphi_{a}\rangle \nonumber\\
&=& \frac{1}{4\pi}\langle \vec{x}_{b}|\frac{\hbar}{\hat{\vec{p}}^{2}/(2m)
- e^{2}/r - E}|\vec{x}_{a}\rangle 
\end{eqnarray}
which establishes the gauge independence of the Green's function for negative 
$E$. Although we here used the same notation for the state $|\xi, \eta, \varphi
\rangle$ in (43) and (53), the meaning of these states are quite different.
This difference is explicitly exhibited in (20) $\sim$ (22). It is important to realize that the exact path integral is performed for the off-shell states in (43).

\section{ Conclusion}
The path integral treatment of a general separable Hamiltonian 
of Liouville-type is formulated on the basis of the Jacobi's principle of least action by using a gauge theoretical technique, and it has been illustrated for the hydrogen atom in parabolic coordinates.

I thank C. Bernido, P. Cvitanovic, J. Klauder, L. D. Faddeev, V. Sayakanit,  and H. Sugawara for helpful comments.

\end{document}